\documentclass[twocolumn, aps,pra, superscriptaddress, 10pt]{revtex4-2}

\usepackage[T1]{fontenc} 
\usepackage[utf8]{inputenc}

\usepackage{float}

\usepackage{graphicx} 
\graphicspath{ {./figures/} }

\usepackage{siunitx}
\usepackage{braket}
\usepackage{amsfonts, amssymb, amsmath}
\usepackage{bbold}
\usepackage{hyperref}
\usepackage[capitalize]{cleveref}
\usepackage{bbold}
\usepackage{xspace}
\usepackage{overpic}
\usepackage{mathtools}
\usepackage{booktabs}

\newcommand{\maleate}{(1-${}^{13}\mathrm{C}$,$\mathrm{d}_6$)-dimethyl maleate \xspace}
\newcommand{\DMAD}{(1-${}^{13}\mathrm{C}$,$\mathrm{d}_6$)-dimethyl acetylenedicarboxylate \xspace}
\newcommand{\Cth}{${}^{13}\mathrm{C}$\xspace}

\begin{document}

\title{Towards a unified picture of polarization transfer - pulsed DNP and chemically equivalent PHIP}

\author{Martin C. Korzeczek}
\affiliation{Institute of Theoretical Physics and IQST, Albert-Einstein Allee 11, Ulm University, 89081 Ulm, Germany}
\author{Laurynas Dagys}
\affiliation{NVision Imaging Technologies GmbH, 89081 Ulm, Germany}
\author{Christoph M{\"u}ller}
\affiliation{NVision Imaging Technologies GmbH, 89081 Ulm, Germany}
\author{Benedikt Tratzmiller}
\affiliation{Institute of Theoretical Physics and IQST, Albert-Einstein Allee 11, Ulm University, 89081 Ulm, Germany}
\affiliation{Carl Zeiss MultiSEM GmbH, 73447 Oberkochen, Germany}
\author{Alon Salhov}
\affiliation{NVision Imaging Technologies GmbH, 89081 Ulm, Germany}
\affiliation{Racah Institute of Physics, The Hebrew University of Jerusalem, Jerusalem, 91904, Givat Ram, Israel}
\author{Tim Eichhorn}
\affiliation{NVision Imaging Technologies GmbH, 89081 Ulm, Germany}
\author{Jochen Scheuer}
\affiliation{NVision Imaging Technologies GmbH, 89081 Ulm, Germany}
\author{Stephan Knecht}
\affiliation{NVision Imaging Technologies GmbH, 89081 Ulm, Germany}
\author{Martin B. Plenio}
\thanks{M.B.P and I.S. contributed equally to this work and its leadership. Correspondence should be addressed to martin.plenio@uni-ulm.de or ilai@nvision-imaging.com}
\affiliation{Institute of Theoretical Physics and IQST, Albert-Einstein Allee 11, Ulm University, 89081 Ulm, Germany}
\author{Ilai Schwartz}
\thanks{M.B.P and I.S. contributed equally to this work and its leadership. Correspondence should be addressed to martin.plenio@uni-ulm.de or ilai@nvision-imaging.com}
\affiliation{NVision Imaging Technologies GmbH, 89081 Ulm, Germany}


\begin{abstract}
Nuclear spin hyperpolarization techniques, such as dynamic nuclear polarization (DNP) and parahydrogen-induced polarization (PHIP), have revolutionized nuclear magnetic resonance and magnetic resonance imaging. In these methods, a readily available source of high spin order, either electron spins in DNP or singlet states in hydrogen for PHIP, is brought into close proximity with nuclear spin targets, enabling efficient transfer of spin order under external quantum control. Despite vast disparities in energy scales and interaction mechanisms between electron spins in DNP and nuclear singlet states in PHIP, a pseudo-spin formalism allows us to establish an intriguing equivalence. As a result, the important low-field polarization transfer regime of PHIP can be mapped onto an analogous system equivalent to pulsed-DNP. This establishes a correspondence between key polarization transfer sequences in PHIP and DNP, facilitating the transfer of sequence development concepts. This promises fresh insights and significant cross-pollination between DNP and PHIP polarization sequence developers.
\end{abstract}

\maketitle

\setlength{\parindent}{0pt}

\newcommand*{\Cthirteen}{$^{13}$C}

\section{Introduction} Nuclear magnetic resonance (NMR) and 
magnetic resonance imaging (MRI) applications are severely 
limited by low detection sensitivities. These are rooted in
the weak thermal equilibrium nuclear spin polarization of the
detection targets at room temperature which, typically, amounts to a few parts 
per million (ppm) per Tesla of applied magnetic field. Achievable
magnetic field strengths in magnetic resonance (MR) devices
are approaching their physical limits and significant future 
increases in the sensitivities of detection coils appear 
challenging. This places stringent limits on the minimally 
detectable quantities under MR and thus presents a major
obstacle towards the goal of metabolic imaging applications
at physiologically relevant metabolite concentrations, which,
in turn, could have a major impact on cancer treatment 
\cite{kurhanewiczAnalysisCancerMetabolism2011,kurhanewicz2019hyperpolarized,wang2019hyperpolarized}.

Therefore, MR-based metabolic imaging requires an increase in the MR signal 
by orders of magnitude and has motivated the search for methods 
that can deliver such gains. The last decades have seen
a wide variety of proposals to overcome the MR detection challenge 
by increasing the nuclear spin polarization beyond its thermal
equilibrium value - thus achieving hyperpolarized nuclei. Promising
approaches include parahydrogen-induced polarization 
(PHIP)~\cite{BowersW1986,BowersW1987,PravicaW1988} by catalytic
addition \cite{ReineriBA2015} or by reversible exchange \cite{AdamsAA+2009}, 
dynamic nuclear polarization (DNP) at low temperatures \cite{ArdenkjaerFG+2003} 
and by optically polarized persistent \cite{LondonSC+2013} and
non-persistent \cite{EichhornTS+2013,EichhornPJ+2022} electron 
spins. These methods have the demonstrated ability 
of achieving
nuclear spin polarization in the percent range which enables MR
signal increases of around four orders of magnitude. 

\begin{figure*}
    \centering
    \includegraphics[width=0.95\textwidth]{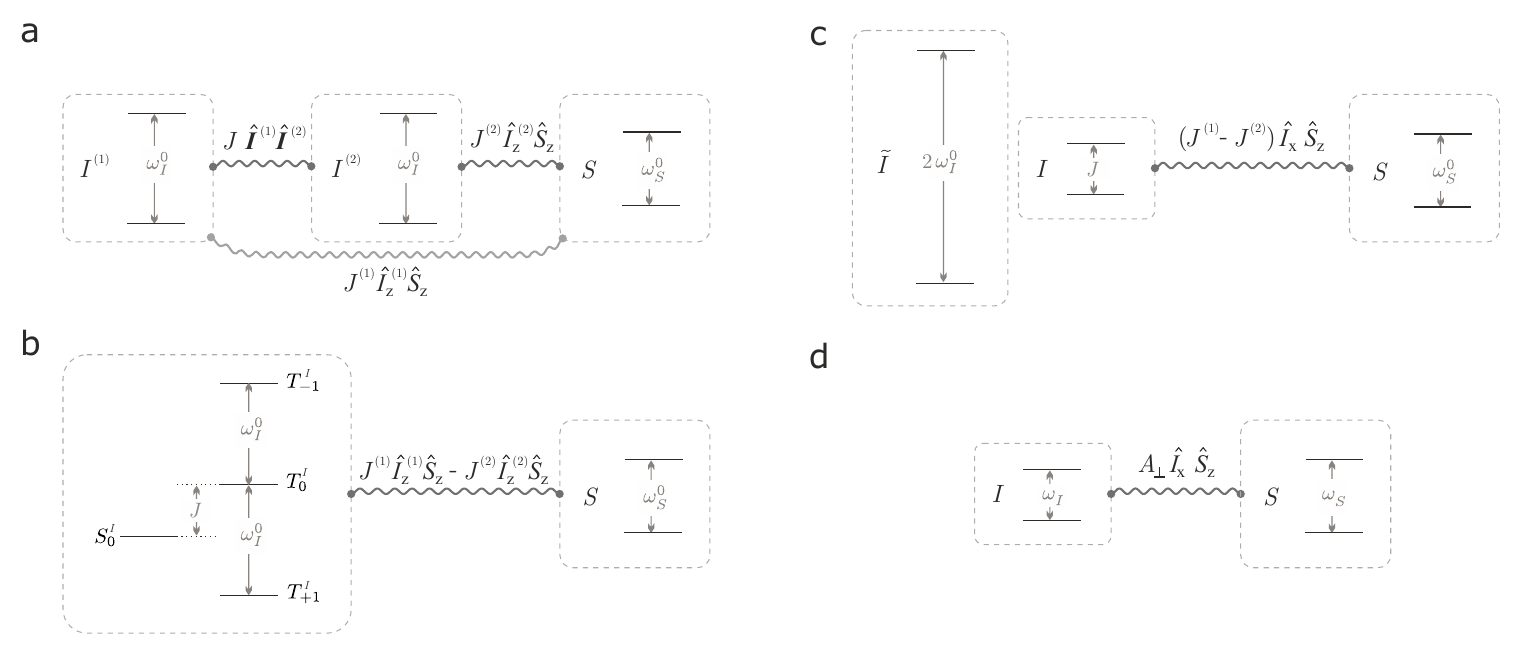}
    \caption{Depiction of the spin system, energy levels and relevant Hamiltonian terms, demonstrating the mapping between DNP and PHIP. (a) the system of two hydrogen spins $I^{(1)}$, $I^{(2)}$ and one heteronuclear spin $S$. (b) The same system with the hydrogen spins $I^{(1)}$, $I^{(2)}$ in the singlet-triplet basis notation, as typically done in PHIP. (c) decomposing the hydrogen energy states into the interacting ($I$) and non-interacting ($\tilde{I}$) pseudospins. (d) Hamiltonian of nuclear spin and electron spin in typical DNP systems. From (c), (d) it becomes evident that the pseudospin formalism for PHIP has significant similarities to a DNP system.}
    \label{fig:pol-hamiltonian}
\end{figure*}

Despite significant differences in the underlying physical systems and
experimental detail these methods share important conceptual similarities. 
Indeed, their common motif is the combination of three key 
ingredients. First, a readily available source of polarization albeit in a form 
that is not directly usable for imaging such as thermally or optically 
polarized electron spins or singlet order in parahydrogen; secondly, 
the ability to bring these spins into close proximity with the target 
nuclei, such as \Cthirteen, that need to be hyperpolarized and, thirdly, 
control methods based on externally applied fields that enable robust
and efficient transfer of polarization from source to target.

Each of the proposed hyperpolarization methodologies have developed their 
own control protocols and languages in which to describe them. The different 
energy scales between DNP (electron to nuclear spin transfer, three orders 
of magnitude difference in energy levels) and PHIP (nuclear spin to nuclear 
spin transfer) have resulted in seemingly distinct constraints and demands 
on the control methods, keeping development in these fields separated. The realization of
potential commonalities between the fields would benefit considerably from
the availability of "dictionaries" and "recipes" that facilitate the translation between
control protocols.

In this work we will expose the commonality between two important classes of 
PHIP and DNP control protocols for polarization transfer. We use the framework 
of \cite{Eills2017} in the near-equivalence regime for chemically equivalent 
PHIP molecules (i.e. the chemical shift between the two proton spins is negligible 
compared to their J-coupling), where we define a pseudospin 
for the strongly coupled protons which has an energy separation equal to the 
J coupling. In this framework, a heteronuclear spin such as \Cthirteen\, has 
an energy splitting that is orders of magnitude larger than that of the pseudospin, 
leading to a regime similar to that encountered in pulsed DNP systems, where 
inter-electron interactions can be neglected (see Fig. \ref{fig:pol-hamiltonian}
for a schematics). This pseudospin takes the role of nuclear spin in DNP while
the \Cthirteen\, takes the role of the DNP electron spin. We demonstrate 
the mathematical equivalence of the descriptions of the two systems and exemplify 
at the hand of explicit examples that this enables the direct translation of a
large number of control protocols from one field to the other. 

We then proceed by concentrating on two practical examples: (1) the 
PulsePol family of  protocols \cite{Schwartz2018} that has been developed 
theoretically and demonstrated experimentally in the framework of DNP with
color centers in diamond to facilitate polarization transfer that is highly 
robust to a broad range of control errors that include detuning errors due
to $B_0$ field inhomogeneities 
and temporal and spatial $B_1$-field fluctuations of external driving 
fields. We demonstrate by numerical simulation that the beneficial features
of robustness translate to the PHIP setting and proceed with a experimental 
work to verify that the predicted efficiency and robustness 
of this new control sequence in the realm of PHIP translates into high levels
of polarization of \Cth  in dimethyl maleate. (2) The recently introduced adiabatic 
amplitude sweeps in the context of PHIP \cite{Marshall2022} was shown to be 
robust and efficient in the regime where the coupling 
still allows for a sweep duration much shorter than the relevant coherence times. 

We demonstrate 
the applicability of this protocol for DNP with optically polarized electrons 
in pentacene molecules inside a naphthalene crystal and the resulting efficiency
advantage over NOVEL, a standard method for polarization transfer in such systems.

\section{Theory}
\label{seq:th_pol_transfer_PHIP}

In this section we derive a mapping between the key physical degrees of freedom 
underlying pulsed dynamical nuclear polarization and those of parahydrogen induced 
polarization for heteronuclear magnetization using chemically equivalent systems. 
We identify the effective qubit degrees of freedom and show that 
the Hamiltonians in both settings can be identified by a suitable basis transformation. 
Armed with this equivalence, we proceed to reveal and describe the parallels between 
various polarization schemes commonly used in PHIP and in DNP, 
thereby exemplifying general translation rules between DNP and PHIP control protocols in the corresponding subfields.

\subsection{Scope of the equivalence}
The equivalence described in this work uses an effective 2-spin description for both
DNP and PHIP. For DNP this limits the scope to systems whose electron linewidth is 
narrow compared to the nuclear Larmor frequency, and for which the solid effect or 
pulsed-DNP schemes are viable \cite{malyDynamicNuclearPolarization2008}. Additionally, 
it is assumed that we only drive the spin $S$ which corresponds to the electron spin 
in DNP and to the heteronuclear spin in PHIP. For PHIP, we limit the system to the case 
of two chemically equivalent proton spins, which will be reduced to a single pseudospin 
in our description, and one heteronuclear spin. This is the typical case for low-field 
PHIP, but is also encountered in high-field PHIP for the case of symmetric molecules 
\cite{Stevanato2017}, as well as signal-amplification by reversible exchange (SABRE) 
polarization of $^{13}C$ \cite{Schmidt2023Sabre} and direct $^{15}N$ polarization 
\cite{royDirectEnhancementNitrogen152017}. Explicitly excluded parahydrogen-based 
systems are high-field PHIP of chemically inequivalent protons and PHIP or SABRE at
ultra-low magnetic fields where the Larmor frequency is on the same order as the
J-coupling (e.g. as in SABRE-SHEATH~\cite{truong201515n}). 

An additional related field is the transfer of population between the singlet and 
magnetized states of two hydrogen spins, either utilizing long-lived  singlet 
states \cite{fengAccessingLonglivedNuclear2012,carravettaT1Limit2004,carravettaLongLivedNuclearSpin2004} 
or para-hydrogen \cite{pileioStorageNuclearMagnetization2010}. Notwithstanding the lack 
of an equivalence on the level of the Hamiltonian, for many \Cthirteen\, polarization sequences 
it has been demonstrated that the same sequences applied to the hydrogens can still be used: 
In \cite{DeVience2013} SLIC was originally developed for this situation, and similarly S2M, 
introduced in \cite{pileioStorageNuclearMagnetization2010}, is deeply related to the S2hM 
sequence \cite{Stevanato2017} which is discussed in this work. More recently, in 
\cite{sabba2022symmetry}, PulsePol has been shown to also be applicable for achieving 
hydrogen-magnetization in PHIP and for singlet-NMR.

\subsection{Polarization transfer in pulsed DNP} \label{sec:theoryDNP}

Dynamical nuclear polarization has the goal of mediating the transfer of polarization from a source spin ($S$), typically of large energy splitting such as an electron, to a target spin ($I$) of lower energy splitting, typically a nuclear spin (see Fig.~\ref{fig:pol-hamiltonian} where the level splittings of $I$ and $S$ are denoted by $\omega_I$ and $\omega_S$ respectively). We regard the case in which pulses can be applied to spin $S$ to facilitate direct transfer of polarization between the two spins. Here, the spins are coupled via the Hamiltonian term $H_{SI}=A_\perp \hat{I}_x \hat{S}_z$. In the case of NV$^-$ centers in diamond, for example, $S$ is given by a two-level subsystem of the electronic ground-state triplet \cite{Tratzmiller2021}. Furthermore, the spin $S$, with gyromagnetic ratio $\gamma_S$, may be driven by a time-dependent magnetic field $B_1(t)$ whose Rabi frequency $\Omega=|\gamma_S B_1|$ may reach or even exceed significantly $\omega_I$. Typically, the energy scales are assumed to fulfill the conditions
\begin{align}\label{eq:scale_hierarchy}
    \omega_S\gg\ \gamma_S |B_1|, \omega_I\ \gg A_\perp.
\end{align}
The corresponding Hamiltonian reads ($\hbar=1$)
\begin{align} \label{eq:pol-transfer-H}
    H = \omega_I \hat{I}_z + \omega_S \hat{S}_z + A_\perp \hat{I}_x  \hat{S}_z + \gamma_S B_1(t)\hat{S}_x. 
\end{align}
Under the assumptions of eq. \ref{eq:scale_hierarchy} and a suitable choice of $B_1(t)$, pulsed or continuous wave, an application of Average Hamiltonian Theory \cite{Choi2020} 
then results in the effective Hamiltonian 
\begin{align} \label{eq:pol-transfer-H-eff} 
H^{\text{eff}} = H_{SI}^{\text{eff}}=\dfrac{A_\ast}{2} (\hat{I}_x \hat{S}_x + \hat{I}_y \hat{S}_y).
\end{align}
in the frame corotating with the influence of both magnetic fields as defined by $\omega_I \hat{I}_z+\omega_{B1} \hat{S}_z + \gamma_S B_1(t) \hat{S}_x$ where $\omega_{B1}$ is the frequency of the driving field $B_1(t)$. In this work all Hamiltonians marked with "eff" refer to this frame. 
Here, the effective coupling $A_\ast \le A_\perp$, a function of the chosen drive $B_1(t)$, effects coherent polarization transfer between the two spin degrees of freedom $I$ and $S$. A full derivation for the case of SLIC/NOVEL can be found in SI \ref{seq:SI-AHT}. In order to achieve this type of Hamiltonian, for {\em pulsed} polarization schemes, the waiting times between pulses typically need to scale with a single free parameter $\tau$ which has to be matched to $\omega_I$ to reach resonance and achieve a non-zero $A_\ast$. For {\em continuous} wave schemes it is the Rabi frequency that typically has to match $\omega_I$.

Important criteria for good polarization schemes are the attained coupling strength $A_\ast/A_\perp$, the scheme's robustness to errors in the driving field such as a resonance-offset ($\Delta$) due to $B_0$-inhomogeneities,   temporal and $B_1$-field fluctuations ($\Omega_{error}$), and also the simplicity of the scheme's implementation in practice.

\subsection{Polarization transfer in PHIP for chemically equivalent molecules and heteronuclear transfer}

We proceed to establish the equivalence of the DNP setting of section A and a typical 
setting for PHIP in the low-field regime with chemical equivalence and transfer to heteronuclear spins. We derive a basis 
transformation such that in the new basis the PHIP system can be written as a set of 
pseudo-spins whose dynamics is governed by a Hamiltonian that takes the form of eq. 
\ref{eq:pol-transfer-H}. A summary of the resulting correspondences between PHIP and DNP is given
in Tab.~\ref{tab:phip-dnp_dictionary}.

\begin{table*}
    \centering 
    \begin{tabular}{lll}
          & \textbf{PHIP} (chemically equivalent, &  \textbf{DNP} (pulsed) \\
          & \phantom{PHIP (} heteronuclear) & \\
         \midrule
         Definitions of & $2\hat{I}_z = \ket{T_0}\bra{T_0} - \ket{S_0}\bra{S_0}$ & $2\hat{I}_z = \ket{\uparrow}\bra{\uparrow} - \ket{\downarrow}\bra{\downarrow}$ \\
         Spin-I operators & $2\hat{I}_x = \ket{S_0}\bra{T_0} + \ket{T_0}\bra{S_0}$ & $2\hat{I}_x = \ket{\downarrow}\bra{\uparrow} + \ket{\uparrow}\bra{\downarrow}$\\
         \midrule
         Spin $I$ & \small $\{\ket{T_0},\ket{S_0}\}$ effective spin & $\{\ket{\downarrow},\ket{\uparrow}\}$ nuclear spin \\
          & $J\ \hat{I}_z$ &  $\omega_I\ \hat{I}_z$   \\[0.2cm]
         Spin $S$ & heteronuclear spin  & electron spin  \\
         & $\omega_S\ \hat{S}_z$ & $\omega_S \hat{S}_z$ \\[0.2cm]
         $I$-$S$ coupling & $ (J^{(1)}-J^{(2)}) \hat{I}_x \hat{S}_z$ &  $A_\perp \hat{I}_x \hat{S}_z$ \\[0.2cm]
         Initial state & $I$ polarized & $S$ polarized \\[0.2cm]
         Scheme: & SLIC \cite{DeVience2013} & NOVEL \cite{Henstra1988}  (\ref{sec:results}\ref{sec:B1sweepPETS}) \\ 
         [0.1cm]
         & S2hM~\cite{Eills2017, Stevanato2017} &  NV nuc. spin init.~\cite{Taminiau2014}\\[0.1cm]
         & PulsePol (this work, \ref{sec:results}.\ref{sec:PulsePolforPHIP}) & PulsePol~\cite{Schwartz2018} \\  [0.1cm]
         & ADAPT~\cite{Stevanato2017b} (\ref{sec:results}.\ref{sec:PulsePolforPHIP}) & TOP-DNP~\cite{Tan2019} \\ [0.1cm]
         & adiab. $B_1$ sweeps~\cite{Marshall2022} (\ref{sec:results}.\ref{sec:PulsePolforPHIP}) & RA-NOVEL~\cite{Can2017} (\ref{sec:results}\ref{sec:B1sweepPETS}) \\ 
         \bottomrule
    \end{tabular}
    \caption{A tabular correspondence for polarization schemes in the regarded subfields of PHIP and DNP. The full names of the polarization schemes are Spin-Lock Induced Crossing for SLIC, Nuclear Spin Orientation via Electron Spin Locking for NOVEL, Singlet to Heteronuclear Magnetization Transfer for S2hM, NV nuclear spin initialization, Alternating Delays Achieve Polarization Transfer for ADAPT, Time-Optimized Pulsed DNP for TOP-DNP, adiabatic $B_1$ sweeps and Ramped-amplitude NOVEL, for RA-NOVEL.}
    \label{tab:phip-dnp_dictionary}
\end{table*}

To this end, we consider a system consisting of two hydrogen nuclear spins $I^{(1)}$ and $I^{(2)}$ and a carbon nuclear spin $S$, all spin-$1/2$, which are all coupled via J-coupling (cf.~Fig.~\ref{fig:pol-hamiltonian}a). Following \cite{Eills2017}, the corresponding Hamiltonian is given by 
\begin{align}\label{eq:PHIP_bare}
H &= \omega^0_I \hat{I}^{(1)}_z +\omega^0_I \hat{I}^{(2)}_z +\omega_S \hat{S}_z \\ &\quad + J\,\hat{\vec{I}}^{(1)}\cdot\hat{\vec{I}}^{(2)} + J^{(1)} \hat{S}_z \hat{I}^{(1)}_z+J^{(2)} \hat{S}_z \hat{I}^{(2)}_z,\nonumber
\end{align}
where $\omega^0_I$ ($\omega_S$) is the hydrogen (carbon) Larmor frequency, $J$ is the inter-hydrogen $J$-coupling strength, and $J^{(1)}$, $J^{(2)}$ are the heteronuclear coupling strengths between the respective hydrogen spin and the carbon spin. 
Here, all J-couplings are defined as angular frequencies.
The singlet-triplet basis
\begin{eqnarray*}
    \ket{S_0} &:=&(\ket{\uparrow\downarrow}-\ket{\downarrow\uparrow})/\sqrt{2}\\
    \ket{T_{-}}&:=&\ket{\downarrow\downarrow}\\ \ket{T_0}&:=&(\ket{\uparrow\downarrow}+\ket{\downarrow\uparrow})/\sqrt{2}\\ 
    \ket{T_{+}}&:=&\ket{\uparrow\uparrow}
\end{eqnarray*}
in the Hilbert space of hydrogen nuclei, that satisfies 
\begin{eqnarray*}
    \vec{I}^{(1)}\cdot\vec{I}^{(2)} \ket{S_0} &=& -\frac{3}{4}\ket{S_0}, \\
    \vec{I}^{(1)}\cdot\vec{I}^{(2)}\ket{T_{\alpha}} &=& +\frac{1}{4}\ket{T_{\alpha}}
\end{eqnarray*}
is the eigenbasis of the inter-hydrogen $J$-coupling term. This basis allows us to re-partition 
the hydrogen manifold into the direct sum of two pseudo-spins: We define the first as the
singlet-triplet pseudo-spin $I$ with $2\hat{I}_z := \ket{T_0}\bra{T_0}-\ket{S_0}\bra{S_0}$ 
and the corresponding $2\hat{I}_x := \ket{T_0}\bra{S_0}+\ket{S_0}\bra{T_0}$. The second is
the complementary pseudo-spin $\hat{\tilde{I}}$ with $2\hat{\tilde{I}}_z:= \ket{T_{+}}\bra{T_{+}}
-\ket{T_{-}}\bra{T_{-}}$. If we use the spin labels to denote the respective Hilbert spaces, 
this basis change can be regarded as switching from the the true-spin basis $I^{(1)}\otimes 
I^{(2)}\otimes S$ to the pseudo-spin basis $(I\oplus \tilde{I})\otimes S$ which both span the 
same combined Hilbert space. With this new basis, we can use the identities
\begin{align}
    \hat{I}^{(1)}_z &=\hat{\tilde{I}}_z + \hat{I}_x\\
    \hat{I}^{(2)}_z &=\hat{\tilde{I}}_z - \hat{I}_x\\
    \hat{\vec{I}}^{(1)}\cdot\hat{\vec{I}}^{(2)} &= \hat{\tilde{I}}_z^2 - \hat{I}_z^2 + \hat{I}_z\\
    \hat{\tilde{I}}^2_z &= \frac{1}{4}\hat{P}_{\tilde{I}}
\end{align}

to transform the Hamiltonian. Here, operators such as $\hat{I}_z$ need to be regarded as acting on the 
combined 4-level Hydrogen manifold, and $\hat{P}_{I}$ and $\hat{P}_{\tilde{I}}$ are the corresponding 
projectors onto the basis states spanned by $I$ and $\tilde{I}$. Using these definitions, we arrive at
\begin{align} \label{eq:phip_new_full}
H	&=  J  \hat{I}_z + \omega_S \hat{S}_z +  (J^{(1)}-J^{(2)}) \hat{I}_x \hat{S}_z \nonumber \\
	&\quad + 2\omega^0_I\hat{\tilde{I}}_z + \frac{J}{4}\hat{P}_{\tilde{I}} - \frac{J}{4} \hat{P}_I + \dfrac{J^{(1)}+J^{(2)}}{2} \hat{P}_{\tilde{I}} \hat{S}_z
\end{align}

The first line corresponds exactly to the terms of 
the DNP Hamiltonian of \eqref{eq:pol-transfer-H}  
albeit without the driving term. The latter can be added to both \eqref{eq:PHIP_bare} and \eqref{eq:phip_new_full} without modifications.  

The second line describes the level-splitting of the third pseudospin $\tilde{I}$ and energy shifts of the $I$ and $\tilde{I}$ sub-spaces, all of which commute with all the terms in the first line. Thus, the PHIP Hamiltonian is indeed equivalent to the DNP setting and thus allows for the mapping of polarization sequences when making the correspondence
\begin{eqnarray}
    J &\leftrightarrow& \omega_I\, ,\\
   J^{(1)}-J^{(2)} &\leftrightarrow& A_\perp\, ,\\
   \omega_{B1} = \omega_S & \leftrightarrow & \omega_{B1} = \omega_S.
\end{eqnarray}
The energy hierarchy condition \eqref{eq:scale_hierarchy} 
implies that we need to be in the near-equivalence regime \cite{Eills2017} 
where $J\gg J^{(1)}-J^{(2)}$ for the sequences regarded in this work to be applicable as they rely on average Hamiltonian theory and thus $A_\perp/\omega_I\ll 1$. The full translation table is given in Table~\ref{tab:phip-dnp_dictionary}.

\section{Results} \label{sec:results}
\subsection{Comparing polarization schemes}  

A wide variety of polarization schemes with different advantages and drawbacks have been 
proposed in the literature, both for PHIP in chemically equivalent systems and for pulsed DNP. Here, we aim to give an overview
of the underlying mechanism and main properties of a number of commonly used polarization 
schemes in each setting. For each scheme we emphasize the similarity and differences between 
the counterparts in DNP or PHIP. In practice, robustness properties and transfer schemes 
are relevant also in situations where the Hamiltonian of eq.~\ref{eq:pol-transfer-H} does
not apply or additional unwanted contributions are present. The reader is referred to 
\cite{Choi2020,Levitt2008,Tratzmiller2021} for treatments that expand on the scope of the 
sequence properties discussed in this work. 
 
The largest difference between the regarded PHIP and the DNP settings is the initial state of the two 
(pseudo-)spins: Whereas in PHIP it is the pseudo-spin $I$, not subject to external control, 
which is initially polarized (given by the singlet-state), in DNP it is the electronic spin 
$S$, subject to external control, which starts out in a polarized state. For polarization schemes, 
this is of importance for the specific choice of the initial or final pulses in the polarization 
sequences that are driving the spin $S$: In principle, 
most schemes need both initial and final pulses to ensure a polarization exchange along $\hat{S}_z$ 
and $\hat{I}_z$ in the lab frame. This is due to the central part of all the schemes discussed here
which creates a related, but different effective interaction of the form $H_{SI}^{\text{eff}}=(A_\ast/2)(\hat{I}_x \hat{S}_z 
+ \hat{I}_y \hat{S}_y )$. 
This interaction transfers polarization along $\hat{S}_x$ and $\hat{I}_z$ axes. The additional initial and final $\pi/2$ pulses now ensure that the effective interaction takes the intended form of \eqref{eq:pol-transfer-H-eff}. If there is no 
initial polarization on spin $S$ (PHIP), the initial pulse has no effect and becomes optional. 
Correspondingly, for an initially polarized spin $S$ (DNP), the state of $S$ after polarization 
transfer is unpolarized, such that any final pulse becomes optional. In the figures representing 
the polarization schemes, we always include both the initial and the final pulse, which makes the
schemes applicable for both the PHIP and DNP setting. In practice, of course, it will often be 
expedient to remove the respective optional pulse.

For pulsed schemes, any stated values for the resonance condition for $\tau$ refers to the limit $A_\ast/\omega_I\rightarrow 0$. This condition will experience corrections that scale in the ratio of $A_\ast/\omega_I$\cite{Haeberlen1968}.

\subsubsection{SLIC and NOVEL}

The conceptually simplest polarization scheme drives the spin $S$ with a constant Rabi frequency 
that matches the level-splitting of $I$, i.e. $\Omega(t) = \omega_I$ in $\gamma_s B_1(t)=\Omega(t)\,
\cos(\omega_S t)$, which results in a polarization transfer at the maximally possible rate $A_\ast 
= A_\perp$ (cf. eq. \ref{eq:pol-transfer-H-eff}). In the field of PHIP, this scheme is known as
Spin-Locking Induced Crossing (SLIC) \cite{DeVience2013} (cf. \cite{Eills2017} for an application of SLIC to heteronuclear transfer), in the field of DNP as Nuclear spin 
Orientation via Electron spin Locking (NOVEL) \cite{Henstra1988} (see \cite{LondonSC+2013} for
an implementation and \cite{cai2013large} for a numerical analysis based on NV centers) and is known 
as Hartmann-Hahn 
resonance in nuclear spin double resonance experiments \cite{HartmannH62}. If the constant intensity 
drive is of phase $X$, by itself it would transfer polarization along the $\hat{S}_x$ and $\hat{I}_z$ 
axes. Thus, suitable $\pi/2$ pulses are needed to rotate any initial (final) polarization along 
$\hat{S}_x$ from (to) $\hat{S}_z$ as indicated in Fig.~\ref{fig:seqs}a.

{\em Discussion of Robustness Properties --} Limitations of this method are that the acceptable 
error in the Rabi frequency $\Omega_{error}$ is proportional to $A_\ast$, while the acceptable resonance 
offset $\Delta$ is proportional to $\omega_I$. The continuous wave drive is easy to implement
in experiment, however accurately matching the Rabi frequency $\Omega(t)$ to $\omega_I$ can be difficult due
to magnetic field fluctuations and power fluctuations of the source. The latter may be addressed 
to some extent by concatenation schemes which add suitably chosen sidebands to the driving field 
that protect against power fluctuations \cite{cai2012robust}.

\subsubsection{S2hM and NV nuclear spin initialization}

Trains of $n$ regularly spaced $\pi$ pulses are commonly used to dynamically decouple the 
driven spin from its environment or to make it susceptible to a specific resonant frequency. 

In our case, an interpulse temporal separation of $\tau = \pi/\omega_I$ creates the effective 
interaction Hamiltonian 
\begin{align} 
H^{\text{eff}}_{SI,1}= A_\ast \hat{I}_x \hat{S}_z,
\end{align}
with $A_\ast = (2/\pi)\ A_\perp \approx 0.64 \ A_\perp$. 

A second train of $\pi$ pulses can induce
\begin{align} 
H^{\text{eff}}_{SI,2}= A_\ast \hat{I}_y \hat{S}_y 
\end{align}
with a suitable transition:
Shifting $\hat{S}_z\rightarrow \hat{S}_y$ can be achieved with a $\pi/2$ pulse of phase $X$, and a waiting time of $(\pi/2)/\omega_I = \tau/2$ induces $\hat{I}_x\rightarrow \hat{I}_y$.

If the total duration of the two pulse trains is shorter than the polarization transfer ($2n\tau \ll (2\pi)/A_\ast$), the effective Hamiltonians $H^{\text{eff}}_{SI,1}$ and $H^{\text{eff}}_{SI,2}$ averages to  $H_{SI}^{\text{eff}}=(A_\ast/2) (\hat{I}_x \hat{S}_z + \hat{I}_y \hat{S}_y)$.
As in the case of SLIC and NOVEL, suitable initial or final pulses can now be added in order to reach $\hat{S}_z$-$\hat{I}_z$-transfer (cf.~Fig.~\ref{fig:seqs}c,h).

In Singlet to heteronuclear Magnetization transfer (S2hM) ~\cite{Eills2017,Stevanato2017}, $n$ is chosen such that the polarization is transferred already after the second pulse train. As in this regime Hamiltonian averaging cannot be used to reach the transfer Hamiltonian eq. \ref{eq:pol-transfer-H-eff} exactly, the reachable polarization is slightly below 100~\% (cf.~Table~1~in~\cite{Stevanato2017}).
For DNP, an almost identical scheme has been used for NV nuclear spin initialization~\cite{Taminiau2014}. 

The robustness properties of S2hM can be significantly improved with the use of phase cycling~\cite{Souza2012,Genov2017} and well-chosen pulse timing to refocus the effect of $\Delta$ for all waiting times using the $\pi$-pulses. Due to the variable number of $\pi$-pulses per train, this creates the need to use different phase cycles for different system parameters $A_\perp$ and $\omega_I$. This can be avoided by using a relatively small and fixed $n$ for each $\pi$-train and appending the $\pi/2$-pulse and the $\pi/(2\omega_I)$ waiting time after each of a variable number of $\pi$-trains.
The pulse trains are now repeated until polarization is fully transferred. Fig.~\ref{fig:seqs}c,h) shows one such choice and a detailed description as well as the robustness properties of S2hM without phase cycling can be found in the S.I.\,.
This approach has been used in~\cite{Schwartz2018} as motivation for developing the even more robust PulsePol sequence, which will be described in the next section.

{\em Discussion of Robustness Properties --} For sequences based on pulse trains, the acceptable error in the resonance offset $\Delta$ is proportional to the Rabi frequency $\Omega$. The acceptable errors on the Rabi frequency typically remain tied to $A_\ast$, but the use of error-correcting phase-cycles or composite pulses~\cite{tyckoCompositePulsesPhase1985,levittCompositePulses1986,Stevanato2017} can improve this to a scaling with the Rabi frequency $\Omega$ itself.
Compared to SLIC and NOVEL, the improved robustness properties come at the cost of moderately reduced speed.

\subsubsection{PulsePol}

\begin{figure*}[t]
    \centering
    \includegraphics[width=\linewidth]{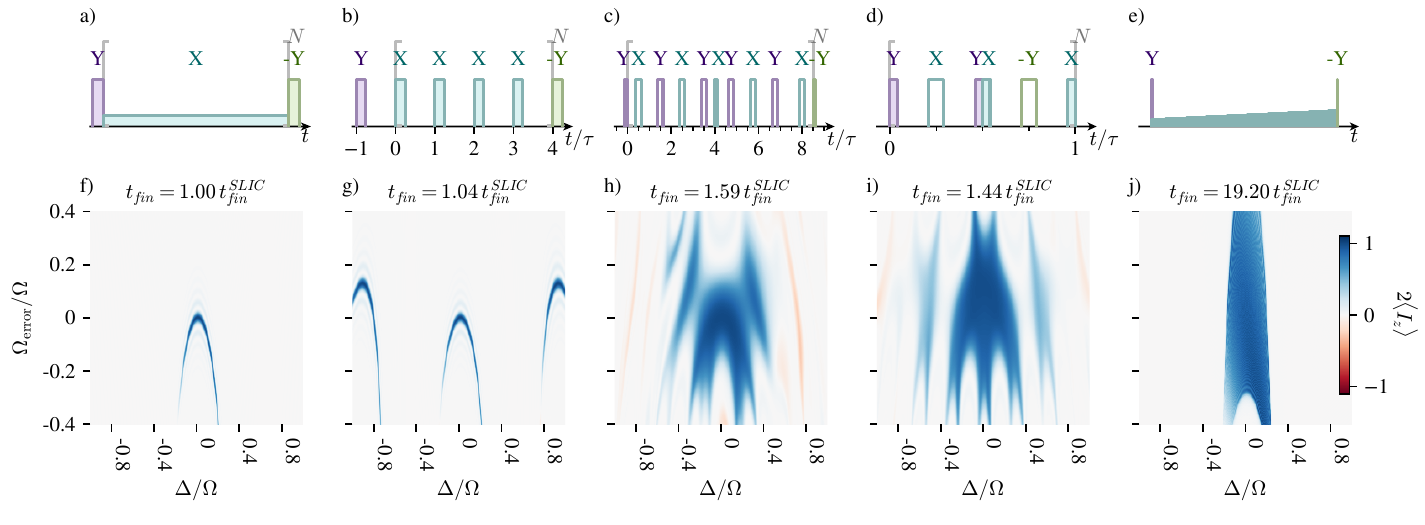}
    \caption{Drawings of the different polarization schemes introduced in the results section (a,b,c,d,e) and respective heatmaps showing the numerical results for the transferred polarization for varying B0- and B1-errors at time $t_{fin}$ (f,g,h,i,j). The parameters used in the simulation are $\Omega=100\,A_\perp$ and $\omega_I=24\,A_\perp$. In the sequence plots, $\pi/2$ pulses and continuous-wave pulses are filled, whereas $\pi$ pulses are not.
    a,f) show the SLIC sequence (NOVEL in DNP),
    b,g) show the S2hM sequence in the case of PHIP and the NV-nuclear spin initialization in DNP with the slight adjustment of using a fixed n=8 XY8 phase cycle and the waiting time distributed equally before the first and fourth $\pi$ pulse after the $\pi/2$ pulse. The corresponding results for S2hM using fixed phases are given in the S.I.
    c,h) show PulsePol,
    d,i) show the ADAPT sequence in the case of PHIP and TOP-DNP in the case of DNP. The chosen pulse duration corresponds to a $\pi/2$ pulse and the resonance condition equivalent to SLIC/NOVEL was used.
    e,j) show a linear $B_1$ sweep with $\Omega\in[0.6\omega_I,\ 1.4\omega_I]$. The duration of the swept pulse in (e) is not to scale.}
    \label{fig:seqs}
\end{figure*}

PulsePol is a polarization sequence developed for DNP in \cite{Schwartz2018}.  It consists of a phase-shifted pair of three pulses each, with a central $\pi$ pulse surrounded by symmetric waiting times, and $\pi/2$ pulses at both ends (cf.~Fig~\ref{fig:seqs}d,i). 
With regards to how the effective Hamiltonian arises, PulsePol can be considered as a minimally short variation of a S2hM-type sequence as described in the previous section: In PulsePol, each pulse train consists of only a single pulse. A pair of $\pi/2$ pulses between each pulse train reorients the current $S$ polarization along $\pm \hat{S}_z$  at their intersection, and a shift $\hat{I}_x\rightarrow -\hat{I}_y$ between pulse trains is reached by uniformly reducing all inter-pulse delays $\tau$ instead of inserting an additional waiting time of $\pi/(2\omega_I)$. The uniform pulse delays strengthen the refocusing properties and the changed waiting time improves the effective coupling by roughly $14\,\%$ compared to S2hM. 
The robustness properties of PulsePol become clear if we consider that the composition of a $\pi$ pulse surrounded by the two 90 degree phase-shifted $\pi/2$ pulses creates the effect of a composite $\pi$ pulse \cite{levittCompositePulses1986}, while the intended effective Hamiltonian is created during the waiting times as described above. Additionally, every set of two repetitions corresponds to the $XY4$-sequence \cite{maudsleyModifiedCarrPurcellMeiboomGillSequence1986} based on this composite pulse which provides a second layer of error correction.

With a total duration $\tau$ for one repetition, the optimal resonance is given by $\tau=3\pi/\omega_I$, which results in $A_\ast = [2\,(2+\sqrt{2})/(3\pi)] A_\perp\approx 0.72\,A_\perp$.

{\em Discussion of Robustness Properties --} In PulsePol, both the acceptable Rabi amplitude error and 
resonance offset scale with the Rabi frequency $\Omega$. Together with its high degree of robustness, 
the slightly increased speed of transfer compared to S2hM makes PulsePol a very attractive sequence. 
However, in setups with a limited bandwidth it can be challenging to implement the consecutive $\pi/2$
pulses. This problem can be alleviated by PulsePol variations which include a waiting time in between
the $\pi/2$ pulses \cite{Tratzmiller2021}.

\subsubsection{ADAPT and TOP-DNP}

Both Time-optimized DNP (TOP-DNP)~\cite{Tan2019} and Alternating Delays Achieve Polarization Transfer (ADAPT)~\cite{Stevanato2017b} are families of sequences that can be described as pulsed versions of 
SLIC or NOVEL: By exchanging the CW pulse with $\Omega=\omega_I$ with an alternating sequence of short 
pulses and waiting times, the same effective interaction can be approximated while improved robustness 
properties and additional freedom in the choice of the waiting time are gained. In \cite{Tan2019}, the 
latter is used to achieve polarization transfer also in the regime $\Omega <\omega_I$, which however 
comes at the cost of lowering the speed of polarization transfer ($A_\ast$). A similar approach is also 
used in~\cite{Casanova2019} to enable transfer in the $\Omega < \omega_I$ regime which can lead to an
overall reduction in average pulse energy.

Using the 'natural' resonance where $\langle\Omega(t)\rangle_t=\omega_I$, the coupling strength can approach 
that of SLIC/NOVEL $A_\ast \approx A_\perp$ when using small rotation angles in each pulse. Fig.~\ref{fig:seqs}g,h 
use this condition together with $\pi/2$ pulses as a representative example of the properties of this scheme 
in the $\Omega>\omega_I$ regime. 

{\em Discussion of Robustness Properties --} As visible in the numerical results shown in Fig.~\ref{fig:seqs}h, the robustness to resonance-offset $\Delta$ gains side-bands with successful polarization transfer in a region scaling with $\Omega$ where each band has robustness properties very similar to SLIC or NOVEL. Thus, in each band, acceptable Rabi errors scale with $A_\ast$ and the acceptable resonance offset scales with $\omega_I$. 
As $\Omega$ defines the achievable control strength, it provides an upper bound for the scaling behavior of robustness in any sequence. Thus, the limitation of the robustness to resonance offset to values $\Delta<\omega$ stops being of relevance in the regime of $\Omega < \omega$.

\subsubsection{adiabatic $B_1$ sweeps}

Swept polarization methods~\cite{Can2017,Kozinenko2019,Rodin2021,Marshall2022,Eichhorn2014}, can only partially be described with the effective Hamiltonian picture used in this work: They can  usually be regarded as a variation of SLIC or NOVEL in which one of the parameters such as the drive amplitude $\Omega$ relevant for the resonance is swept at a well-chosen speed. In words, the condition is that for every value of the swept parameter, resonance to spins with this value is approximately fulfilled for an amount of time that allows polarization to be transferred.
However, the effect on the driven spin itself is now more complicated. As there are no repetitions in the sequence, not only the induced effective interaction Hamiltonian, but also the state evolution of the driven spin can be non-trivial. Often, slow adiabatic sweeps are used to achieve a well-defined evolution. 
Schemes which sweep the resonance offset $\Delta$ include the integrated solid effect \cite{Eichhorn2014} for DNP. 

Fig.~\ref{fig:seqs}e,j show the sequence and robustness for an adiabatic $B_1$ sweep similar to~\cite{Marshall2022} (PHIP) and \cite{Can2017} (DNP). For simplicity, the sweep rate is chosen as $\dot{\Omega}=A_\perp^2/(2\pi)$. Compared to SLIC and NOVEL (Fig.~\ref{fig:seqs}a,f), the sweep significantly enhances the robustness to amplitude errors. However this increased robustness comes at the cost of correspondingly increasing the duration of the sequence. In a similar fashion, frequency-swept pulses can reach very high bandwidths with successful polarization transfer.

{\em Discussion of Robustness Properties --} If the total duration of the scheme is not a limiting factor, swept polarization methods can provide very high robustness to errors in the swept parameter, while often being simpler to implement than pulsed schemes.

\subsection{Experimental translation between DNP and PHIP}
We can now utilize the correspondence between PHIP in chemically equivalent systems and pulsed DNP as illustrated in Table~\ref{tab:phip-dnp_dictionary} to apply the most fitting polarization transfer sequences to our experimental conditions. We consider two experimental systems - polarization of $^{1}H$ spins in naphthalene using triplet-DNP and achieving high polarization in the $^{13}C$ spin of dimethyl maleate using PHIP.
In the former, using an amplitude sweep allows for a significant increase in the transferred polarization without significantly more complicated $B_1$ drives. The latter demonstrates the applicability of the highly robust PulsePol to PHIP.

\begin{figure}[t]
    \centering
    \includegraphics[width=0.45\textwidth]{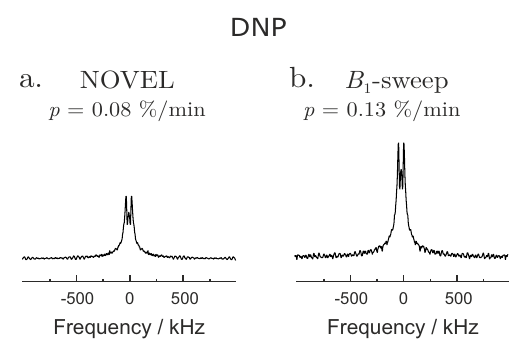}
     \caption{Comparison of the achieved polarization rates in a pentacene doped naphthalene crystal when applying the NOVEL and the adiabatic amplitude sweep. The spectra were taken after polarizing for 20\,s. We see a significant increase in polarization rates for the amplitude sweep, which will allow higher final polarization values.}
    \label{fig:PETS-sweeps}
\end{figure}

\subsubsection{adiabatic B1 sweeps with PETS} \label{sec:B1sweepPETS}

Photo excited triplet states (PETS) can be used as a tool for DNP, where the optically induced polarization of the PETS is transferred to close by nuclear spins. Here, a pentacene doped naphthalene crystal is used, where the pentacene acts as PETS and the protons in naphthalene are the nuclear spins for which macroscopic polarization is desired. Various sequences for this polarization transfer are known and have been shown to work, e.g. NOVEL and the integrated solid effect (ISE) \cite{HenstraDW1988,HenstraLSW1990,Eichhorn2014}.

The experimental setup is described in~\cite{Marshall2021}. We performed experiments at a B$_0$ field of 175\,mT, corresponding to a hydrogen Larmor frequency of around 7.5\,MHz.
Using NOVEL, successful polarization transfer was achieved (cf.~Fig.~\ref{fig:PETS-sweeps}(a)), although the presence of amplitude errors proved to be a limiting factor: Using a linear adiabatic $B_1$-sweep/RA-NOVEL, the transfer rate over a duration of 20\,s could be improved by up to 60\,\% when using an optimized sweep rate (cf.~Fig.~\ref{fig:PETS-sweeps}(b)).

\begin{figure}[t]
    \centering
    \includegraphics[width=0.45\textwidth]{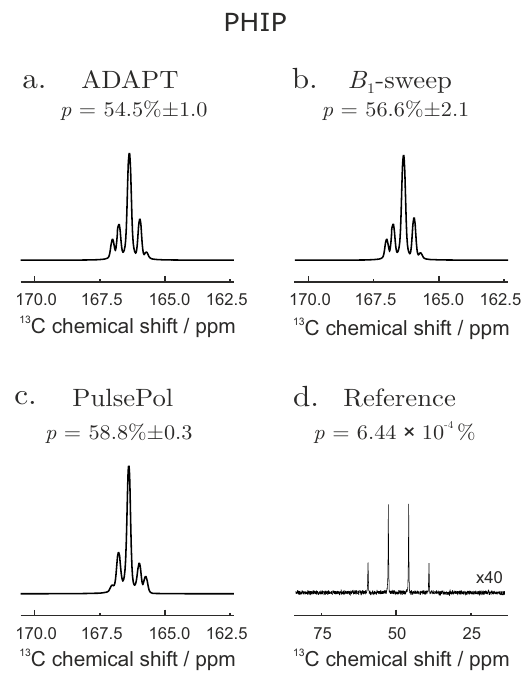}
     \caption{Hyperpolarized \Cth spectra of 15~mM \maleate acquired at 1.8~T, obtained by 
     using different polarization transfer methods operating at 100~$\mathrm{\mu}$T bias field. 
     (a) Transfer induced by \Cth-ADAPT consisting of 5  [90$^\circ$-$\mathrm{\tau}$] loops with 
     free evolution time of 21.4~ms. (b) adiabatic $B_1$ sweep with transverse field amplitude 
     swept from 0 to 25~Hz in 2 seconds. (c) Application of a single PulsePol echo block with free 
     evolution time set to 20.5~ms. (d) Conventional \Cth NMR spectrum obtained on a $^{13}$C-methanol 
     at thermal equilibrium sample (averaged over 4 transients). Estimated \Cth polarization levels 
     are indicated.}
    \label{fig:PHIP-results}
\end{figure}

\subsubsection{Pulsepol for PHIP} \label{sec:PulsePolforPHIP}
\maleate provides a PHIP system with strong coupling $A_\perp$ which strains the condition of near-equivalence \eqref{eq:scale_hierarchy}. Nonetheless, polarization sequences can still be applied successfully while the strong coupling allows for a correspondingly increased robustness to detuning or amplitude errors (cf.~Fig.~\ref{fig:SI_app_1}) which in our case is sufficient to reach high absolute polarization for all of the regarded sequences ADAPT, adiabatic $B_1$ sweeps and PulsePol.
We used solutions of 
20~mM \DMAD and 5~mM [Rh(dppb)(COD)]BF$_4$ dissolved in acetone-d$_6$. 
For each experiment, hydrogenation of the precursor solutions at a pressure of 10~bar for 2~s was applied with a subsequent purging with nitrogen gas for 18~s at room temperature. The 92\%-\textit{para}-enriched hydrogen was produced at 21~K by an Advanced Research Systems generator. To sustain singlet-order during this process, 
continuous-wave decoupling of $^1$H was applied.

The polarization schemes were applied at a bias field of $B_0=100\,\mu$T (Fig.~\ref{fig:PHIP-results}) and a drive amplitude of $\Omega= (2\pi)$ 50~Hz for pulses. The adiabatic $B_1$ sweep 
in Fig.\ref{fig:PHIP-results}(b) involved a transverse field amplitude sweep from 0~Hz to 25~Hz performed in 2 seconds followed by a 90 degrees pulse. 

After the polarization transfer, the level of \Cth polarization was assessed by rapidly inserting the sample into a Bruker benchtop 1.8~T NMR magnet and acquiring the \Cth NMR spectrum after the excitation by a single pulse. 
The polarization was estimated by comparing spectra against a reference spectrum of thermally polarized neat \Cth-methanol. 
For the absolute polarization levels, the concentration of the product was calculated by recording $^1$H spectra of both the product solutions and an aqueous solution of 492~mM DMSO.
The error margins in Fig.~\ref{fig:PHIP-results} were established by repeating each experimental procedure three times. The highly similar polarization levels reached indicate that all three schemes provide sufficient robustness to cover the bulk of the distribution of detuning and amplitude errors in the system, such that the strong robustness of PulsePol provides only a small advantage.

\section{Discussion}
Many insights in physics can be obtained by the translation of tools and techniques from 
one field to another. In the same sense, the direct translation between the near-equivalence 
regime in chemically equivalent PHIP and pulsed DNP enables a cross-pollination between the fields. First, this straightforwardly 
provides new insights for existing control protocols. For example, the case of long MW pulses 
studied thoroughly in TOP-DNP can shed light on the comparable case of long RF pulses in the 
ADAPT protocol, as will be encountered for example in implementations of ADAPT at very low 
magnetic fields. Second, this translation allowed us to identify sequences developed for one 
physical system and apply them in the other system, where they could exceed the performance of the 
current state of the art. This includes sequences recently developed in \cite{wiliDesigningBroadbandPulsed2022}, the use of PulsePol for PHIP polarization transfer, and 
the use of adiabatic $B_1$ sweeps for the transfer of polarization from optically polarised 
electron spins to nuclear spins in pentacene:naphthalene crystals.

Another interesting area of application would be in the refinement of polarization transfer in SABRE, where polarization in the near-equivalence regime has been recently shown to lead to superior enhancements of $^{13}C$ nuclear spins in partially deuterated molecules \cite{Barskiy2019,Schmidt2023Sabre}.

We expect that additional insights can be gained from the above equivalence, not only in the field 
of polarization transfer but other instances where a similar pseudospin formalism is used. For 
example, methods and ideas from singlet-NMR could be applied to quantum information processing 
and nanoscale sensing using NV centers in diamond and vice-versa.

\section*{Acknowledgements}
The authors acknowledge financial support by the German Federal Ministry of Education and Research (BMBF) under the funding program quantum technologies - from basic research to market via the project QuE-MRT (FKZ: 13N16447). 
MBP and MK additionally acknowledge financial support by the ERC Synergy grant HyperQ (grant no. 856432), and the 
Center for Integrated Quantum Science and Technology (IQST).

\section*{Declaration of competing interests}
The authors declare the following competing financial interest(s): Laurynas Dagys, Tim R. Eichhorn, Stefan Knecht, Christoph M{\"u}ller, Alon Salhov, and Ilai Schwartz are employees of NVision Imaging Technologies GmbH (NVision). Martin B. Plenio and Ilai Schwartz are co-founders of NVision. NVision is commercializing products in the field of magnetic resonance hyperpolarization.

\bibliography{references}

\clearpage
\appendix

\subsection*{Details regarding the effective Hamiltonian reached via Average Hamiltonian Theory} \label{seq:SI-AHT}
In \ref{sec:theoryDNP}, we referred to a "suitable choice" for $B_1(t)$ to reach \eqref{eq:pol-transfer-H-eff} from \eqref{eq:pol-transfer-H} by applying Average Hamiltonian Theory.
Here, we provide a complete calculation using the SLIC sequence as an example.

In the lab frame, the system is described by the Hamiltonian from \eqref{eq:pol-transfer-H}:
\begin{align}
    H(t) = \omega_I \hat{I}_z + \omega_S \hat{S}_z + A_\perp \hat{I}_x  \hat{S}_z + \gamma_S B_1(t)\hat{S}_x. 
\end{align}
Here, $B_1(t) = 2\Omega(t)/\gamma\ \cos(\omega_{B1} t + \varphi(t))$ describes our linearly polarized driving field which oscillates with angular frequency $\omega_{B1}=\omega_S-\Delta$ where the resonance offset $\Delta$ is usually unintended.

We can now enter the frame corotating with the Larmor precessions (except for $\Delta$) using $H^{\text{L}}=\omega_{B1} \hat{S}_z + \omega_I \hat{I}_z$ and the corresponding unitary $U^{\text{L}} = e^{-i \omega_{B1} t \hat{S}_z} e^{-i \omega_{I} t \hat{I}_z} $ and reach
\begin{align}
    H^L(t) &\approx  \Delta \hat{S}_z + A_\perp (\hat{I}_x \cos(\omega_I t) - \hat{I}_y \sin(\omega_I t))  \hat{S}_z \nonumber \\ 
    &+ \Omega(t) (\cos\varphi(t)\hat{S}_x+ \sin\varphi(t)\hat{S}_y),
\end{align}
where we discarded terms oscillating with $2\omega_{S}$ using the scale hierarchy \eqref{eq:scale_hierarchy}. In regimes where $B_0$ is weaker, the use of rotating waves for $B_1$ avoids the need for this approximation.
Finally we can enter the frame corotating with the influence of our drive, also called the toggling frame, as defined by 
\begin{equation}
    U(0)=\mathbb{1}, \ i\partial_t U(t) =  \Omega(t) (\cos\varphi(t)\hat{S}_x+ \sin\varphi(t)\hat{S}_y) + \Delta \hat{S}_z.
\end{equation}
In this frame, only the interaction between the spins remains part of the Hamiltonian
\begin{align}
    H^{\text{tog}}(t) &= A_\perp (\hat{I}_x \cos(\omega_I t) - \hat{I}_y \sin(\omega_I t)) U^\dagger(t) \hat{S}_z U(t) \nonumber \\
    &=:A_\perp (\hat{I}_x \cos(\omega_I t) - \hat{I}_y \sin(\omega_I t)) \hat{S}^{\text{tog}}_z(t).
\end{align}
Here, we used that $U(t)$ commutes with the spin $I$ operators and defined the toggling-frame $\hat{S}_z$ operator $\hat{S}^{\text{tog}}_z(t)$. Thanks to $A_\perp \ll \omega_I$ we can now approximate the Hamiltonian $H^{\text{tog}}(t)$ with its time average over a duration $T\sim 2\pi/\omega_I$ and reach a constant effective Hamiltonian. For this we assume that the chosen drive returns the state of $S$ to its original state at time $T$, i.e. $U(T)=U(0)$, and the same for the Larmor precession on spin $I$, that is $\exp{i\omega_I T} = 1$. This ensures that the same effective Hamiltonian describes the time spans $[0,T]$, $[T,2T]$ et cetera for as long as the same driving sequence is applied. The effective Hamiltonian becomes
\begin{align}
    H^{\text{eff}} &= \dfrac{1}{T}\int_0^T \mathrm{d} t\ H^{\text{tog}}(t) \\
        &= A_\perp\ \dfrac{1}{T}\int_0^T \mathrm{d} t\ (\hat{I}_x \cos(\omega_I t) - \hat{I}_y \sin(\omega_I t)) \hat{S}^{\text{tog}}_z(t). \nonumber
\end{align}
It becomes apparent that $H^{\text {eff}}$ contains an $\hat{I}_x$ component proportional to the frequency-$\omega_I$ cosine contribution of $\hat{S}^{\text{tog}}_z(t)$ and similarly for $\hat{I}_y$ and the $\omega_I$ sine contribution of $\hat{S}^{\text{tog}}_z(t)$.

In case of SLIC, we describe the initial and final pulses as instant at times $t=0$ and $t=NT$ respectively, such that
\begin{align}
    \hat{S}^{\text{tog}}_z(0^-) &= \hat{S}_z \nonumber \\
     \overset{\left(\pi/2\right)_{Y}}{\rightarrow} & \hat{S}^{\text{tog}}_z(0^+) = U^{\text{ini}\dagger}(0^+) \hat{S}_z U^{\text{ini}}(0^+) = - \hat{S}_x \\
   & U^{\text{ini}}(0^+) = \cos(\pi/4)\mathbb{1}+i\sin(\pi/4) 2 \hat{S}_y.
\end{align}
As the duration of this pulse is negligible, we need not regard any corresponding effective Hamiltonian here.

During the spin-locking pulse of SLIC, we have $(\Omega(t),\varphi(t)) = (\omega_I, 0)$, which leads to $U(t)=\exp{(-i\omega_I t \hat{S}_x)} = \cos(\omega_I t/2) \mathbb{1}+ i\sin(\omega_I t/2) 2\hat{S}_x$.
This gives us 
\begin{align}
\hat{S}^{\text{tog}}_z(t) &= U^{\text{ini}\dagger}(0^+) U^\dagger(t) \hat{S}_z U(t) U^{\text{ini}}(0^+) \\
&= U^{\text{ini}\dagger}(0^+) ( \cos(\omega_I t) \hat{S}_z + \sin(\omega_I t) \hat{S}_y ) U^{\text{ini}}(0^+) \nonumber \\
&= -\cos(\omega_I t) \hat{S}_x + \sin(\omega_I t) \hat{S}_y, \nonumber
\end{align}
which together with $T=2\pi/\omega_I$ leads to the effective Hamiltonian
\begin{align}
    H^{\text{eff}}_{SLIC} &= A_\perp\ \dfrac{1}{T}\int_0^T \mathrm{d} t\ (\hat{I}_x \cos(\omega_I t) - \hat{I}_y \sin(\omega_I t)) \hat{S}^{\text{tog}}_z(t) \nonumber \\
    &= A_\perp \left( \dfrac{ -\hat{I}_x \hat{S}_x - \hat{I}_y \hat{S}_y}{2} \right),
\end{align}
which now corresponds to a flip-flop interaction. The opposite sign compared to \eqref{eq:pol-transfer-H-eff} does not change its properties with regards to our purposes.
The final $\pi/2$ pulse is now needed to let the toggling frame coincide with the frame $L$ if the final state of $S$ is of importance.

\subsection*{Details of the Hamiltonian equivalence}
In \cref{seq:th_pol_transfer_PHIP}, we introduce the pseudo-spin basis $(I\oplus \tilde{I})\otimes S$ to arrive at the 
Hamiltonian given in \cref{eq:phip_new_full} which, in turn, is equivalent to the Hamiltonian in the DNP setting if 
restricted to the $I\otimes S$ subspace. Here, we provide the intermediate steps of the required calculations. 

We define the pseudo-spin operators $I_z = (\ket{T_0}\bra{T_0}-\ket{S_0}\bra{S_0})/2$, 
$I_x = (\ket{T_0}\bra{S_0}+\ket{S_0}\bra{T_0})/2$ and $\tilde{I}_z= (\ket{T_{+1}}\bra{T_{+1}}-\ket{T_{-1}}\bra{T_{-1}})/2$. 
For their squares we find

\begin{align}  
	I_z^2 = \dfrac{1}{4}(\ket{T_0}\bra{T_0}+\ket{S_0}\bra{S_0}) =: \frac{1}{4}\hat{P}_{\tilde{I}}
\end{align} 
and
\begin{align} 
	\tilde{I}_z^2 = \dfrac{1}{4}(\ket{T_{+1}}\bra{T_{+1}}+\ket{T_{-1}}\bra{T_{-1}}) =: \frac{1}{4}\hat{P}_I
\end{align}
with the respective projectors $\hat{P}_I$ and $\hat{P}_{\tilde{I}}$ onto subsystems $I$ and $\tilde{I}$ in the 
Hydrogen-manifold. Note that in this pseudo-spin basis, all operators in $I\oplus \tilde{I}$ need to be considered 
as acting on the full 4-level Hydrogen-manifold. In contrast, the original basis $I^{(1)}\otimes I^{(2)}$ allows 
the corresponding operators to be considered as true single-spin operators due to the tensor-product structure.

The $J$-coupling contribution can now be expressed in the new basis, resulting in
\begin{align} 
	\hat{\vec{I}}^{(1)}\cdot\hat{\vec{I}}^{(2)} &= \dfrac{1}{4}(\ket{T_{+1}}\bra{T_{+1}}+\ket{T_{0}}\bra{T_{0}}+\ket{T_{-1}}\bra{T_{-1}}) \nonumber\\
	& \quad - \dfrac{3}{4} \ket{S_{0}}\bra{S_{0}} \nonumber\\
	&= \dfrac{1}{4}(\ket{T_{+1}}\bra{T_{+1}}+\ket{T_{-1}}\bra{T_{-1}}) \nonumber \\
	& \quad -\dfrac{1}{4}(\ket{T_{0}}\bra{T_{0}}+\ket{S_{0}}\bra{S_{0}}) \nonumber\\
	&\quad +\dfrac{1}{2} (\ket{T_{0}}\bra{T_{0}}-\ket{S_{0}}\bra{S_{0}}) \nonumber\\
	&= \frac{1}{4}\hat{P}_{\tilde{I}} - \frac{1}{4}\hat{P}_I + \hat{I}_z.
\end{align}
Finally the $\hat{I}^{(i)}_z$ operators become
\begin{align}
	\hat{I}^{(1)}_z \otimes \hat{\mathbb{1}}^{(2)} &= \dfrac{1}{2} ( \ket{S_{0}}\bra{T_{0}}+\ket{T_{0}}\bra{S_{0}} \nonumber \\
 &\qquad +\ket{T_{+1}}\bra{T_{+1}}-\ket{T_{-1}}\bra{T_{-1}} ) \nonumber \\ 
	&=\hat{I}_x+\hat{\tilde{I}}_z, \\ 
	\hat{\mathbb{1}}^{(1)} \otimes \hat{I}^{(2)}_z &= \dfrac{1}{2} ( -\ket{S_{0}}\bra{T_{0}}-\ket{T_{0}}\bra{S_{0}}  \nonumber \\
 &\qquad +\ket{T_{+1}}\bra{T_{+1}}-\ket{T_{-1}}\bra{T_{-1}} ) \nonumber \\
	&=-\hat{I}_x+\hat{\tilde{I}}_z. 
\end{align}

With these results, we can directly replace all contributions to the Hamiltonian from \cref{eq:PHIP_bare} to reach
\begin{align}
	H &= \omega^0_I \hat{I}^{(1)}_z +\omega^0_I \hat{I}^{(2)}_z +\omega_S \hat{S}_z + J\,\hat{\vec{I}}^{(1)}\cdot\hat{\vec{I}}^{(2)} \nonumber \\ 
	&\quad + J^{(1)} \hat{S}_z \hat{I}^{(1)}_z+J^{(2)} \hat{S}_z \hat{I}^{(2)}_z \nonumber \\
	&= 2\omega^0_I\hat{\tilde{I}}_z + \omega_S \hat{S}_z  + \frac{J}{4}\,(\hat{P}_{\tilde{I}} - \hat{P}_I + 4\hat{I}_z) \nonumber \\
	&\quad + \frac{1}{2}(J^{(1)}+J^{(2)}) S_z \hat{P}_{\tilde{I}} + (J^{(1)}-J^{(2)}) S_z I_x \nonumber \\
	&=  J  \hat{I}_z + \omega_S \hat{S}_z +  (J^{(1)}-J^{(2)}) \hat{S}_z \hat{I}_x  \nonumber \\
	&\quad + 2\omega^0_I\hat{\tilde{I}}_z + \frac{J}{4}\hat{P}_{\tilde{I}} - \frac{J}{4}\hat{P}_I + (\dfrac{J^{(1)}+J^{(2)}}{2}) \hat{P}_{\tilde{I}} \hat{S}_z,
\end{align}
where all terms in the second line commute with all terms in the first line and the former becomes equivalent to \cref{eq:pol-transfer-H} if restricted to the $I\otimes S$ subspace.

\subsection*{Details of the numerical simulation}

The numerical simulations in this work use the pseudo-spin basis $I \otimes S$ for the Hamiltonian. The drive is 
parametrized via a time-dependent amplitude $\Omega(t)$ and phase $\varphi(t)$:
\begin{align*}
	H(t)&= \omega_I \hat{I}_z +A_\perp\hat{S}_z\hat{I}_x \\ 
	&\quad + (1+\dfrac{\Omega_{error}}{\Omega})\cdot \Omega(t)\left( \cos(\varphi(t)) \hat{S}_x + \sin(\varphi(t))\hat{S}_y \right) \\ &\quad + \Delta\hat{S}_z.
\end{align*}
Here, $\Omega_{error}/\Omega$ is the relative error in the Rabi frequency created by the drive on spin $S$, 
and $\Delta$ parametrizes resonance-offset errors of the drive. The system state starts in the state
$$\rho = |+\frac{1}{2}\rangle\langle +\frac{1}{2}|\otimes \frac{1}{2}\hat{\mathbb{1}}$$ where $\hat{I}_z|+\frac{1}{2}\rangle=\frac{1}{2}|+\frac{1}{2}\rangle$. 

The integration in time of the dynamics is governed by the piece-wise constant Hamiltonian $H(t)$. It proceeds by direct exponentiation of the Hamiltonian in each interval where it is constant and chooses, where 
applicable, the number of sequence repetitions $N$ such that the first maximum of the $S$-magnetization is 
reached for the error-free case of $\Omega_{error}=\Delta=0$. For schemes which do not rely on sequence
repetition such as amplitude sweeps and the non-repeating variant of S2hM, $N=1$ is chosen whereas the
duration of the sequence, i.e. the sweep duration or the lengths of the pulse trains, is given by the theoretically optimal values. For the robustness-plots, this value of $N$ is used for calculating the polarization 
reached for a variety of errors in $\Omega_{error}/\Omega$ and $\Delta/\Omega$ and the final polarization is
calculated via the expectation value $\langle S_z\rangle$.

The timing of pulses is calculated in two steps: First, the total duration of a sequence iteration is calculated 
as a sequence-dependent multiple of the Larmor period $\tau$. Secondly, the individual pulses of rotation angle 
$\alpha_i$ and durations $T_i = \alpha_i/\Omega_i$ are distributed in the time interval such that pulses begin 
and end together with their corresponding section defined by $\tau$. Unless specified differently by the sequence, 
the pulses have equal interpulse waiting times to optimize refocusing of rotations induced by $\Delta$. Unless specified otherwise by the polarization
scheme, the maximum Rabi frequency $\Omega(t)=\Omega_i =\Omega$ is assumed to be used for the duration of each 
pulse $i$.

\subsection*{Description of robustness properties}
The intended effective Hamiltonian for polarization transfer between the spins $S$ and $I$ is given by 
\cref{eq:pol-transfer-H-eff} and contains only the Larmor precession of the two spins together with a 
flip-flop contribution of strength $A_\ast \le A_\perp$.

For a full polarization transfer, it is necessary that any accumulating errors which alter the spin state 
remain small over the whole duration $T=2\pi/A_\ast$. For non-error-correcting sequences, unwanted 
Hamiltonian contributions such as a resonance offset $\Delta\hat{S}_z$ will accumulate at an error rate 
$\alpha=\Delta$, whereas first order error-correction due to a drive of Rabi frequency $\Omega$ can suppress 
this to $\alpha \propto (\Delta/\Omega)^2\cdot \omega_I$ where $(\Delta/\Omega)^2$ estimates the scaling 
behavior of the error accumulated during a single repetition of the error-correcting part of the sequence 
and $\omega_I\propto 1/\tau$ estimates the duration of that sequence part.

Thus, for an error of strength $\beta$ and a non-correcting sequence, successful polarization transfer is assured for 
\begin{align}
	\beta = \alpha \ll A_\ast,
\end{align}

and in the case of a (first-order) robust sequence for
\begin{align}
	  c\,(\beta/\Omega)^2\cdot \omega_I = \alpha \ll A_\ast \Leftrightarrow \beta \ll c\, \Omega \sqrt{\dfrac{A_\ast}{\omega_I}},
\end{align}
where $c$ is a proportionality constant depending on the sequence and the details of the type of the error. 
In the main text, we refer to the former as "maximum acceptable error scaling with $A_\perp$" and to the 
latter as "maximum acceptable error scaling with $\Omega$". Note that the latter case still includes an 
additional square root scaling with $A_\perp$.

The errors we regard in this work are detuning errors $H_{err}=\Delta\hat{S}_z$ of the driving field with 
$\beta=\Delta$, and Rabi errors 
\begin{align}
    H_{err}(t)&=\dfrac{\Omega_{error}}{\Omega}\cdot \Omega(t)\left( \cos\varphi(t) \hat{S}_x + \sin\varphi(t)\hat{S}_y \right) 
\end{align}
with $\beta = \Omega_{error}$.

\subsection*{Details of the regarded polarization schemes}

Here, we provide further details about the sequences and robustness properties of the schemes under consideration 
in their respective parameter regimes.

\begin{figure*}
   	\centering
   	\includegraphics[width=\linewidth]{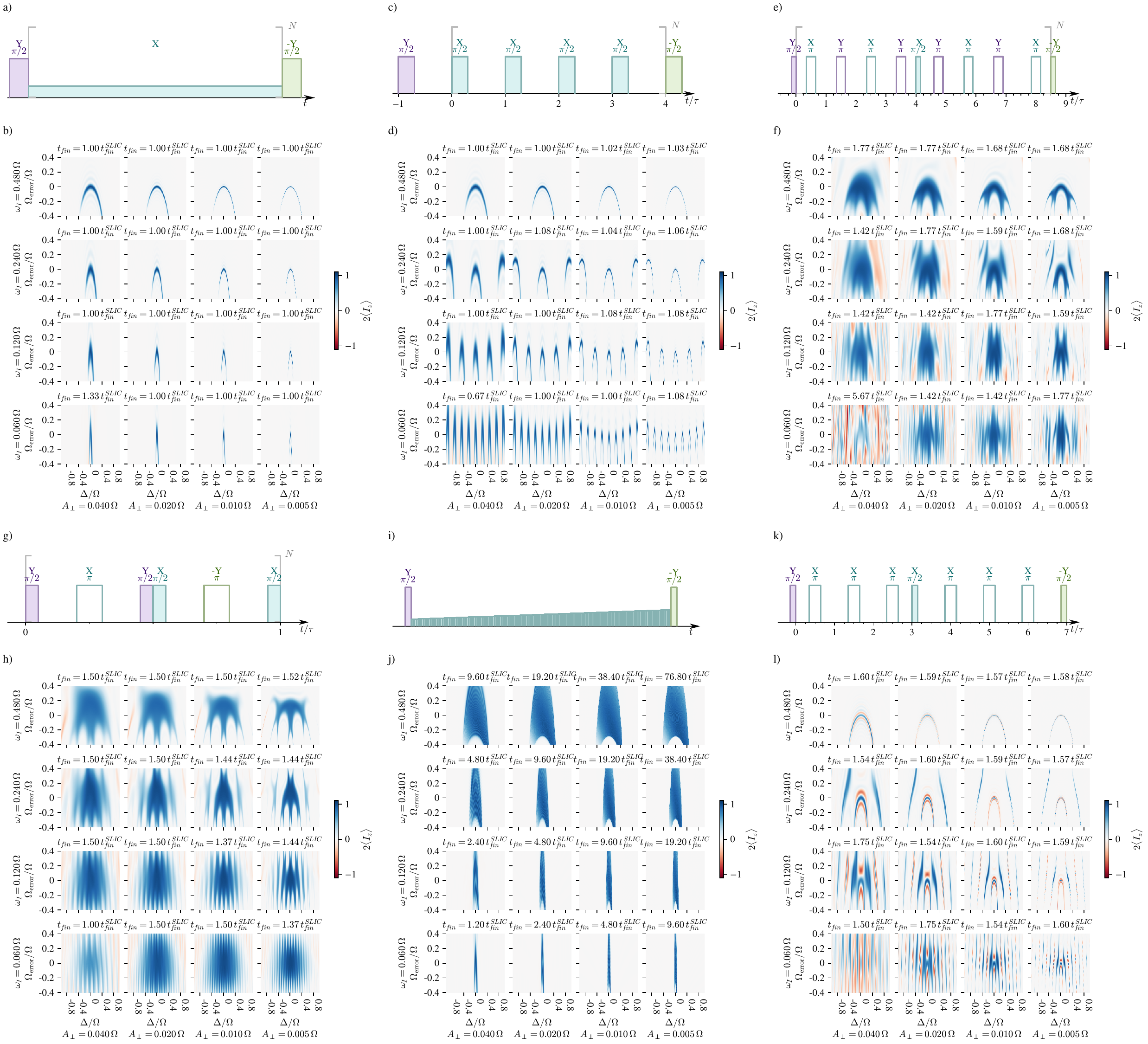}
   	\caption{Detailed robustness plots for all sequences. The sequences are shown in (a,c,e,g,i,k) and the corresponding robustness plots in different regimes of $A_\perp, \omega_I, \Omega$ are shown in (b,d,f,h,j,l). In the robustness plots, from left to right $A_\perp$ and from top to bottom $\omega_I$ is decreased by a factor of 2 in each step and each heatmap shows the success of polarization transfer over varying detuning $\Delta$ and amplitude errors $\Omega_{error}$.
    a,b) show the SLIC sequence (NOVEL in DNP),
    c,d) show the ADAPT sequence in the case of PHIP and TOP-DNP in the case of DNP. The chosen pulse duration corresponds to a $\pi/2$ pulse and the resonance condition equivalent to SLIC/NOVEL was used.
    e,f) show the S2hM sequence with a fixed n=8 XY8 phase cycle and the waiting time distributed equally before the first and fourth $\pi$ pulse after the $\pi/2$ pulse. The corresponding results for S2hM with fixed phases are shown in k,l).
    g,h) show PulsePol,
    i,j) show a linear $B_1$ sweep with $\Omega\in[0.6\omega_I,\ 1.4\omega_I]$. The duration of the swept pulse in (i) is not to scale.
    k,l) show the S2hM sequence with without phase cycling. The corresponding results for S2hM with XY8 phases are given in the e,f).
    }
    \label{fig:SI_app_1}
\end{figure*}

The more general robustness plots in Fig.~\ref{fig:SI_app_1} show the robustness properties of the sequences for 
different regimes of the scale hierarchy $\Omega,\omega_I\gg A_\perp$. From left to right, the coupling between 
the spins $A_\perp$ decreases; from top to bottom $\omega_I$ decreases. 
All results shown remain in the regime $\Omega>\omega_I$. Still, one observes that different sequences display comparative 
advantages in different parameter regimes. All of the robustness plots show the duration $t_{fin}$ of the sequence
compared to a reference given by the ideal application of SLIC $t_{fin}^{SLIC}=2\pi/A_\perp$. Note that these times 
are not necessarily the optimal choices as the algorithm for choosing $N$ is quite simple, and additionally for 
SLIC/NOVEL and ADAPT/TOP-DNP the duration for a single repetition was chosen to correspond to a total rotation angle 
of $2\pi$ although the sequences allow for a finer decomposition. Due to this, the resulting values for $t_{fin}$ as 
well as the calculated robustness properties are less representative of the corresponding sequences in the regime 
$A_\perp \approx \omega_I$, i.e. the lower-left corner of Fig.~\ref{fig:SI_app_1}b,d,f,h,j and m.

\paragraph{SLIC and NOVEL.} This scheme (cf. \cref{fig:SI_app_1} a,b) consists of initial and final $\pi/2$ pulses to 
switch from the $+z$-orientation to the $+x$-orientation (and back) and an intermediate spin-locking pulse with 
amplitude $\Omega = \omega_I$ and phase $X$. In order to treat this scheme as a pulsed scheme, we restricted the 
spin-locking pulse to $N$ repetitions of a pulse of rotation angle $2\pi$ (Fig.~\ref{fig:SI_app_1}a). In Fig.~\ref{fig:SI_app_1}b, 
one can clearly see that close to the error-free point of $\Delta=0=\Omega_{error}$, acceptable amplitude errors 
$\Omega_{error}$ are limited by $A_\perp$ , whereas detuning errors $\Delta$ may reach larger values proportional 
to the amplitude of the spin-locking pulse $\omega_I$. As the amplitude of this pulse is fixed by the spin system, 
it is not possible to take advantage of larger possible values for $\Omega$.

\paragraph{ADAPT and TOP-DNP.} For simplicity and comparability, we only consider the versions with $\pi/2$ pulses and 
stay in the regime of $\Omega>\omega_I\gg A_\perp$. For the regime of $\Omega<\omega_I$ or $\omega_I \sim A_\perp$ 
we refer to the original works \cite{Stevanato2017b,Tan2019}. Choosing $\tau= \pi/(2\omega_I)$, each repetition 
of the sequence consists of four $\pi/2$ pulses of phase $X$ with pulse-to-pulse delay $\tau$. Again, initial and final 
pulses are $\pi/2$ pulses with phases $Y$ and $-Y$ and suitable waiting times such that any two subsequent pulses have 
equal waiting times (cf. Fig.~\ref{fig:SI_app_1}c). The robustness properties of this sequence are shown in Fig.~\ref{fig:SI_app_1}d 
and show a central robust region very similar to that of SLIC/NOVEL, however with additional sidebands in $\Delta$ 
repeating over a region proportional to the Rabi amplitude $\Omega$.

\paragraph{S2hM and NV nuclear initialization} In the main text, we presented the properties of an S2hM variant with 
a fixed pulse train length of $n=8$ $\pi$-pulses
and adjusted waiting times in order to enable phase cycling and full rephasing without 
adjustments to the specific values of $\omega_I$ and $A_\perp$. Comparable robustness properties are still achievable with 
a repetition-free version of S2hM (i.e. $N=1$) as long as phase cycling is used and additionally all waiting times are refocussed 
by $\pi$-pulses. S2hM uses the same initial and final $\pi/2$ pulses of phases $Y$ and $-Y$ as SLIC/NOVEL, and the main 
part of the sequence consists of two pulse trains of $n$ $\pi$-pulses, each surrounded by symmetric waiting times of length 
$\tau/2$ (which together include the pulse-duration) and a transitory $\pi/2$ pulse of phase $X$ together with an extra 
waiting time of $\tau/2$. The time scale of $\tau$ is given by $\tau = \pi/\omega_I$. 
In the robust repeating version of S2hM used in the main text, the $\pi$-pulse trains each consist of four pulses, the 
first with phases $X,Y,X,Y$ and the second with phases $Y,X,Y,X$. To achieve refocusing of all waiting times for any 
number of repetitions $N$, the extra waiting time of $\tau/2-\pi/(2\Omega)$ after the transitory $\pi/2$ pulse 
is separated into two equal parts with the second part being added to the waiting time between the third and fourth 
pulse of the second pi train (cf. Fig.~\ref{fig:SI_app_1}c). Note that for a single repetition $N=1$, this sequence is 
fully equivalent to a usual S2hM sequence with $n=4$ $\pi$-pulses per train and slightly unusual waiting times. For more repetitions ($N>1$), the additional pulse trains per repetition each correspond to a single longer pulse train of $n=8$ $\pi$-pulses with $XY8$ phases. As can be seen in 
Fig.~\ref{fig:SI_app_1}f, this choice allows for $\Omega$-scaling of acceptable errors in both $\Omega_{error}$ and $\Delta$.

As an example for the robustness properties of S2hM without phase cycling, we regard a simpler sequence of equal pulse trains which each consist of $n= \lfloor\dfrac{\pi \omega_I}{2 A}_\ast\rceil$ pulses of equal phase $X$, where $\lfloor \cdot \rceil$ rounds to the closest integer. The waiting time after the transitory $\pi/2$ pulse of phase $X$ is chosen such that the delay between the two pulse trains is $\tau/2$. Finally, an equal waiting time is added before the final $\pi/2$ pulse in order to refocus rotations induced by $\Delta$ during the intermediary waiting times at least in the cases where $N$ is an odd integer (cf. Fig.~\ref{fig:SI_app_1}k). As can be seen in Fig.~\ref{fig:SI_app_1}l, this version of S2hM is susceptible to Rabi errors $\Omega_{error}$, and while the robustness to detuning errors $\Delta$ does scale with $\Omega$, it is still significantly weaker than that of the phase cycled version or that of PulsePol.

\paragraph{PulsePol} In PulsePol, the initial and final pulses are already included in the basic blocks that are repeated 
in the sequence. This part consists of two equal blocks which only differ by a phase shift of all pulses by $\varphi=\pi/2$. 
The first block consists of a central $\pi$ pulse of phase $X$, surrounded by two equal waiting times and begins and ends 
with a $\pi/2$ pulse of phase $Y$. The total duration of a block is $\tau/2$ (cf. Fig.~\ref{fig:SI_app_1}g), while the 
optimal resonance condition is $\tau= 3\pi/\omega_I$. As can be seen in Fig.~\ref{fig:SI_app_1}h, PulsePol retains successful 
polarization transfer in a large uniform region scaling with $\Omega$ both for Rabi ($\Omega_{error}$) and detuning ($\Delta$) 
errors. At a slight cost to the speed of the sequence, it is possible to increase the robust region further by using 
$\varphi=\pi/4$ and $\tau=3.5 \pi/\omega_I$ instead \cite{Tratzmiller2021}.

\paragraph{Adiabatic $B_1$ sweeps} Similar to SLIC/NOVEL, the amplitude sweep consists of initial and final $\pi/2$ 
pulses of phases $Y$ and $-Y$. The main part consists of a continuous pulse with linearly increasing amplitude, 
here $\Omega\in [0.6\,\omega_I,\ 1.4\,\omega_I]$. In our case, this pulse is approximated by 150 pulses of constant 
amplitude each (cf. Fig.~\ref{fig:SI_app_1}i). The total duration of the swept pulse is given by 
$(1.4\,\omega_I-0.6\,\omega_I)/\dot\Omega =\frac{(2\pi)}{A_\perp^2} \cdot 0.8\, \omega_I$. The robustness properties are very 
similar to the ones of SLIC/NOVEL except for the significantly improved robustness to amplitude errors (cf. Fig.~\ref{fig:SI_app_1}j).

\end{document}